\documentstyle[12pt]{article}
\begin{document}

\def\be{\begin{equation}}
\def\ee{\end{equation}}
\def\bea{\begin{eqnarray}}
\def\eea{\end{eqnarray}}
\def\act{\tilde{S}[\rho,\beta]}
\def\ti{\tau_{int}}
%%%%%%%%%%%%%%%%%%%%%%%%%%%%%%%%%%%%%%%%%%%%%%%%%%%%%%%%%%%%%%%%%%%%%%%%%%%%%%%%%

\rightline{\normalsize SMI-1/97}
\rightline{\normalsize KANAZAWA 97-01}
\vspace{1cm}
\centerline{\normalsize\bf A TOY MODEL TEST OF A NEW ALGORITHM}
\baselineskip=22pt
\centerline{\normalsize\bf FOR BOZONIZATION OF FERMION DETERMINANTS}
%\baselineskip=22pt
%\centerline{\normalsize\bf  }
\vspace*{0.6cm}
\centerline{\footnotesize
T.D.~BAKEYEV${}^1$, M.I.~POLIKARPOV${}^{2,3}$,
A.A.~SLAVNOV${}^{1,4}$ and
A.I.~VESELOV${}^{2}$}

\baselineskip=13pt
\vspace*{0.3cm}
\centerline{\it ${}^1$Moscow State University, Vorobievy Gory,
Moscow, 117234, Russia}
\centerline{\it ${}^2$ITEP, B.Cheremushkinskaya 25, Moscow,
117259, Russia}
\centerline{\it ${}^3$Department of Physics, Kanazawa
University, Kanazawa 920-11, Japan}
\centerline{\it ${}^4$Steklov Mathematical Institute, Russian
Academy of Sciences, Vavilova 42, Moscow, 117966, Russia}
\vspace*{0.8cm}
\begin{abstract}

 Different aspects of bozonization algorithm proposed in
\cite{Sl1,Sl2} are tested by numerical simulations of a one
dimensional toy model.  \end{abstract} \section{Introduction.}

Computer simulations in lattice models with fermions meet
serious difficulties due to grassmanian nature of fermionic variables.
The method mainly used so far was based on the Hybrid Monte Carlo
algorithm \cite {DK}. Considerable progress was achieved in this direction
but with the present computer facilities it still does not allow efficient
calculations in models with dynamical fermions.

An alternative approach was put forward by M.Lusher \cite {HL1,HL2}, who
proposed to calculate a fermion determinant replacing it by an infinite
series of boson determinants. However at present the efficiency of this
algorithm is comparable with the efficiency of Hybrid Monter Carlo
method mainly due to the large autocorrelation time. (For a recent
review of different approaches to the problem see \cite {K4}).

So at the moment the problem is still open and alternative approaches
to bozonization should be investigated.
  In papers \cite{Sl1,Sl2} a representation of a D-dimensional fermion
determinant as a path integral of a (D+1)-dimensional Hermitean
bozonic action was proposed. This construction provides a new 
algorithm for numerical simulations of lattice QCD with dynamical 
quarks, although it is generally applicable to any model whose action 
is quadratic in Fermi fields and the matrix of quadratic form is 
positive definite. To study various technical aspects of a simulation 
procedure we carry out in the present paper a detailed study of a one
 dimensional toy model. In the original paper \cite{Sl1} bozonization
procedure was based on the constrained bozonic effective action.
Later it was observed \cite{Sl2} that using a freedom in the choice
of effective bozonic action one can get rid off the constraints which
seems to be more convenient for Monte-Carlo simulations. In this
paper we use the unconstrained version.

 \section{Effective bozonic action.}

Let us consider a D-dimensional lattice fermionic model with the action:

\be S_f=a^D\sum_x \overline{\psi}(x)(B^2+m^2)\psi(x) \ee
where $x$ numerates the sites of a D-dimensional Euclidean hypercubic
lattice, and $B^2$ is some positive bounded operator. (We shall assume
also that the operator $B$ is Hermitean, although it is not really
necessary.)

We introduce (D+1)-dimensional bozonic fields $\phi (x,t)$ which have
the same spinorial and internal structure as $\psi(x)$.

The extra coordinate $t$ is defined on the one dimensional chain of
the length $L$ with the lattice spacing $b$:
\be L=Nb \ ;\ 0\le n<N \ee
We assume that $b\parallel B\parallel\ll 1 $.

The determinant of the operator $B^2+m^2$ can be presented as the
following bozonic path integral \cite{Sl1,Sl2}:

\be Det(B^2+m^2)=\int e^{-S_f}D\overline{\psi}D\psi=
\lim_{L\rightarrow\infty, b\rightarrow 0}\int e^{-S_b[\phi,\chi]}
D\phi^* D\phi D\chi^* D\chi\label{1}\ee
where:
\bea &&S_b[\phi,\chi]=a^D\sum_x b\sum_{n=0}^{N-1}\Bigl[-b^{-2}
(\phi_{n+1}^*(x)\phi_n(x)+h.c.-2\phi_n^*(x)\phi_n(x))+\nonumber\\&&
+b^{-1}(\imath\phi_{n+1}^*(x)B\phi_n(x)+h.c)
+\frac{1}{2}(\phi_{n+1}^*(x)B^2\phi_n(x)+h.c)+\nonumber
\\&& +\sqrt{L}\exp\{-mbn\}(\chi^*(x)(m+\imath B)\phi_n(x)+h.c.)\Bigr]
+\frac{a^D L}{2m}\sum_x\chi^*(x)
\chi(x)\label{1a}\eea
and the free boundary conditions in $t$ are imposed:
\be \phi_n=0 \ ;\ n<0 \ ;\ n\ge N\ee
The bozonic D-dimensional fields $\chi(x)$ have the same spinorial
and internal structure as the fields $\psi(x)$.

The bozonic action (\ref{1a}) is a linearized version of the
expression in the exponent of the following integral:

\bea && I=\int \exp\Bigl\{a^D\sum_\alpha b\sum_{n=0}^{N-1}\Bigl[
b^{-2}(\phi_{n+1}^{\alpha *}\exp\{-\imath B_\alpha b\}\phi_n^\alpha
+h.c.-2\phi_n^{\alpha
*}\phi_n^\alpha)-\nonumber\\&&-\sqrt{L}\exp\{-mbn\} (\chi^{\alpha
*}(m+\imath B_\alpha)\phi_n^\alpha+h.c)\Bigr]-\frac{a^D L}{2m}
\sum_\alpha\chi^{\alpha *}\chi^\alpha\Bigr\}D\phi^* D\phi D\chi^*
D\chi\label{2} \eea where instead of $x$-representation we used a
basis formed by the eigenvectors of the operator $B$, $B_\alpha$
being corresponding eigenvalues.

Indeed, taking into account that $b\parallel B\parallel\ll 1$,
we can expand the expression in the exponent of the integrand
(\ref{2}) in a Taylor series. Keeping only the terms, nonvanishing in
the limit $b\rightarrow 0$, we get the expression (\ref{1a}). For a
finite $b$ the difference between (\ref{1}) and (\ref{2}) is of order
$O(b^2\parallel B\parallel^2)$.

By changing variables
\be \phi_n^\alpha\rightarrow\exp\Bigl\{-\imath B_\alpha nb\Bigr\}
\phi_n^\alpha \ \ ;\ \ \phi_n^{\alpha *}\rightarrow\exp\Bigl\{\imath
B^\alpha nb\Bigl\}\phi_n^\alpha \ee
we can rewrite the eq. (\ref{2}) as the gaussian integral over
$\phi$ with the quadratic form which does not depend on $B_\alpha$.
To calculate this integral it is sufficient to find  a stationary point of
the exponent.

For small $b$ the sum over $n$ can be replaced by the integral and
the equations for the stationary point acquire the form:
\be \ddot \phi^\alpha -\sqrt{L}\chi^\alpha (m-\imath B_\alpha)
e^{-(m-\imath B_\alpha)t}=0 \ee
\be \phi^\alpha(0)=\phi^\alpha(L)=0\ee
(One can show that replacing the sum by the integral also introduces
corrections of order $O(b^2\parallel B\parallel^2$.)

The solution of these equations is:
\be \phi^\alpha (t)=\frac{\chi^\alpha \sqrt{L}}{m-\imath B_\alpha}
\Bigl(e^{-(m-\imath B_\alpha)t}+\frac{t}{L}(1-e^{-(m-\imath
B_\alpha )L})-1\Bigr)\ee
Substituting these solutions to the integrand ,we get:
\be \lim_{b\rightarrow 0}I=\int\exp\Bigl\{-\sum_\alpha \frac{
\chi^{\alpha *}\chi^\alpha}{m^2+B_\alpha^2}(1-2e^{-mL}\cos{B_\alpha L}
+e^{-2mL})\Bigr\}D\chi^* D\chi \label{2.9}\ee
Therefore
\be \lim_{b\rightarrow 0,L\rightarrow\infty}I=det(B^2+m^2)\ee

The equality (\ref{1}) is proven. For finite $b$ and $L$ this
equation has to be corrected by the terms:
\be O(b^2\parallel B\parallel^2)+O(e^{-mL})\label{2.10}\ee

In the next section we apply this construction to simulations of a one
dimensional model.

\section{ Numerical simulations for free fermions on the {\bf $9^1$}
lattice.}

Some technical issues of the new algorithm can be observed by performing
numerical simulations on 1D lattice for free fermions. We used the following
action:
\be S_f=\sum_k {\psi^*}_k(-\partial^2+m^2)\psi_k\label{3.1}\ee
where ${\psi^*}$,$\psi$ are anticommuting Grassman variables,
$\partial$ is symmetrical lattice derivative and $m$ is a mass.
The lattice spacing $a$ has been set equal to 1 for convenience.
One can rewrite the operator in the quadratic form of (\ref{3.1})
as follows:
\be -\partial^2+m^2=B^2+m^2\ee
where $B=(\imath\partial)$ is hermitian matrix. So in accordance
with the discussion above the path integral over Grassmanian
variables $\psi^*, \psi$ can be approximated by the path integral
over bozonic fields $\phi, \chi$. The corresponding bozonic theory
with the multiquadratic action $S_b[\phi,\chi]$ (see eq.(\ref{1a})) can be
simulated straightforwardly using local heatbath and overrelaxation
algorithms.

We note that for the model (\ref{3.1}) the action $S_b[\phi,\chi]$ can
be rewritten in the form:
\be S_b[\phi,\chi]=\act +\tilde{S}[\sigma,\gamma]\ee
where $\rho,\beta$ and $\sigma, \gamma$ are real and imaginary parts of the
fields $\phi,\chi$:
\bea &&\phi=\rho +\imath\sigma \nonumber \\ &&\chi=\beta+\imath\gamma \eea
Hence the path integral over the fields $\phi, \chi$ allows the following
factorization:
\be {\cal Z}_b=\int e^{-S_b[\phi,\chi]}D\phi^* D\phi D\chi^* D\chi=
\Bigl[ \int e^{-\act} D\rho D\beta\Bigr]^2\ee
It makes possible to simulate the theory with the real fields
$\rho ,\beta$
and the action $\act$, accelerating the calculations by the factor of 2.
The resulting partition function must be squared.

Let us firstly consider the updating of the field $\rho$. If all field
variables
except $\rho$ at point $(x,t)$ are kept fixed (where $t$ numerates
points in auxiliary dimension), the action assumes the following form:
\be S_b[\rho(x,t)]= A(\rho(x,t)-E(x,t))^2+{\it const}\ee
where $A$ is a positive constant and $E(x,t)$ is an easily calculable
vector. A local update
\be \rho (x,t)\rightarrow \tilde{\rho}(x,t)=\omega E(x,t)+
(1-\omega) \rho (x,t) +\sqrt{\frac{\omega(2-\omega)}{A}}\eta\ee
where $\eta$ is a gaussian random number of unit variance, fulfills
detailed balance for any $0<\omega\le 2$ \cite {A}.

In our test we used hybrid overrelaxation algorithm, which consists
in the mixing of heatbath ($\omega =1$) and overrelaxation ($\omega =2$)
sweeps with a ratio $1:N_{or}$ \cite {WW}. It is believed that this
algorithm has a dynamical critical exponent $z\approx 1$ if $N_{or}$
is proportional to the correlation length $\xi$.

Generally we subdivided the lattice into 2 sublattices, coloring each
site according to the function \[ C(x,t)=(-1)^t\] One can see that
$E(x,t)$ does not depend on the field variables at the same $t$ and
sites of the same colour do not interact with each other.  We updated
one colour after the other.

Analogously, if all field variables except $\beta$ at point $x$ are
fixed, the action assumes the form:  \be
S_b[\beta(x)]=C(\beta(x)-D(x))^2 +{\it const}\ee where $C$ is a
positive constant. Since field $\beta$ is interacting with
$\rho(x,t)$ for all $t$, the computation of the vector $D(x)$
requires an effort proportional to $N$, i.e. a $\beta$ field update
is almost as expensive as the updating of $\rho$ field.

In our implementation, the iteration is made up of one $\rho$ heatbath
sweep, $N_{or}^\rho$ overrelaxation sweeps, one $\beta$ heatbath sweep
and $N_{or}^\beta$ overrelaxation sweeps. After each iteration the following
function was measured:
\be {\cal R}= <\sum_k \psi_k{\psi^*}_k>   \label{3.2}\ee

To measure the function (\ref{3.2}) using the bozonic approximation
of the fermionic theory with the action (\ref{3.1}), one must
rewrite (\ref{3.2}) as a correlation function of $\rho,\beta$
fields with the measure defined by $\act$. Let us denote:
\be Z_f=\int e^{-S_f}D\psi^* D\psi \ee
\be Z_b=\int e^{-\act}D\rho D\beta \ee
In the section 2 it was proved that
\be Z_f= Z_b^2+ O(b^2\parallel B\parallel^2)+O(e^{-mL}) \ee
Then we can derive:
\be {\cal R}=\frac{1}{2mZ_f}\frac{\delta}{\delta m}Z_f\approx
\frac{1}{2mZ_b^2}\frac{\delta}{\delta m}Z_b^2=\frac{1}{mZ_b}
\frac{\delta}{\delta m}Z_b=-\frac{1}{m} <\frac{\delta \act}{\delta
m}>_{\rho,\beta}\label{3.5}\ee
and
\be {\cal R}_b=-\frac{1}{m}<\frac{\delta \act}{\delta
m}>_{\rho,\beta}\label{3.6}\ee
We used the expression (\ref{3.6}) as a
bozonic approximation to the function (\ref{3.2}). One can see
that the accuracy of this approximation is also given by the
expression (\ref{2.10}).

In table 1 the autocorrelation time dependence is displayed for several
updating schemes. Autocorrelation times were measured for the function
(\ref{3.6}) using the method proposed by Socal \cite {S},namely
\be \ti({\cal R}_b)=\frac{1}{2}+\sum_{i=1}^{M}\frac{C(i)}{C(0)}
\label{f6}\ee
with
\be C(i)=\frac{1}{n-i}\sum_{k=1}^{n-i}({\cal R}_k-\overline{{\cal R}})
({\cal R}_{k+i}-\overline{{\cal R}})\ee
where the $n$ is a number of iterations and $M$ chosen so that
$\ti\ll M\ll n$. An estimate for the error of $\ti$ is given by
\be \sigma_{\ti}^2=\frac{2(2M+1)}{n}\ti^2\ee

One can see that overrelaxing $\beta$ field does not decrease autocorrelation
time substantially (it even may increase $\ti$ in units of CPU time).
Contrary, overrelaxing $\rho$ field improves the autocorrelation
behavior. When adding more overrelaxation sweeps for the given sets
of parameters, $\ti$ in CPU units starts to rise again.  \newpage

\vspace{1cm} \begin{center} \begin{tabular}{|l|l|l|} \hline
Updating &${\cal R}_b$ & $\ti ({\cal R}_b)$\\
\hline
Hh & 0.526(10)  &56(11) \\
\hline
HOh & 0.534(7)& 20(3)\\
\hline
HhOo &0.544(9) &22(4)\\
\hline
HOhO &0.546(6) &10(1)\\
\hline
HOhOo&0.546(6) &9(1)\\
\hline
HOhOO&0.544(4) &4.3(3)\\
\hline
HOOhOO&0.543(3) &2.8(2)\\
\hline
HOOOhOO&0.541(3)&2.2(2)\\
\hline
HOOhOOOo&0.541(2)&1.6(1)\\
\hline
\end{tabular}
\end{center}
{\bf Table 1:} {\small Autocorrelation times in [1/iteration] units
on $9^1$ lattice for $m=4,N=100$ and $b=0.015$.
The letters in the first column give the type and order of sweeps used
per iteration, where H is a $\rho$ heatbath, O is a $\rho$ overrelaxation
and h and o are the $\beta$ updates. The exact value of ${\cal R}$
is 0.5457.}
 \vspace{1cm}

Now we discuss the systematic error and autocorrelation
behavior of the algorithm. We measure the function ${\cal R}_b$ (see
(\ref{3.6})) and $\ti ({\cal R}_b)$ for different sets of parameters
$b$ and $N$.  As it is demonstrated in section 2, the systematic
error is given by:  \be \Delta = \Delta_1
+\Delta_2,\ee where:  \be \Delta_1=Cb^2+Fb^3+O(b^4),\label{3.3s}\ee
\be \Delta_2 =D e^{-mbN},\label{3.3}\ee and $C,F,D$ are some
constants.

To control the systematic error one can fix the parameter $mbN$ and
perform calculations for different values of $b$. We choose
$m=4\ ,\ mbN=8\ $. The
results are shown on Fig.1 where the function ${\cal R}_b(mb)$
(\ref{3.6}) is plotted (the statistical errors are
small). The horizontal line corresponds to the theoretical value
(\ref{3.2}). It is seen that the results converge to
the theoretical value as $mb$ decreases. The fit of expression
(\ref{3.3s}) for $\Delta_1$ gives $C\approx 17$ and $F\approx -43$.
The systematic error $O (b^2)$ is partially compensated by the
$O (b^3)$ error.

On Fig.2 the dependence of the autocorrelation time for
${\cal R}_b$ against $mb$ is plotted. In these and all further measurements
HOhOO scheme is used. From these data we get that
$\ti({\cal R}_b)\approx 0.25/mb$ when $mb\rightarrow 0$.

Now let us fix the parameter $b=0.01$ making $\Delta_1$ less than
$0.002$ and measure ${\cal R}_b$ and $\ti({\cal R}_b)$
for the different number of points in auxiliary dimension $N$.
On Fig.3 the dependence of ${\cal R}_b$ against $N$ is shown.
The horizontal line denotes the theoretical value of (\ref{3.2}).
From these data we get $D<2.5$.

On Fig.4 the dependence of $\ti({\cal R}_b)$ against $N$ is plotted.
The autocorrelation time grows quadratically when $N$ increases, but
the proportionality factor is very small: \newline $\sim 10^{-4}$.
For large $N$ the autocorrelation time can be decreased by adding more
overrelaxation sweeps of the $\rho$ field.  To investigate the
autocorrelation behavior of the algorithm for the cases of practical
importance one needs to study the models of larger dimensionality
with the gauge fields involved.

In conclusion we investigate the slowing down of the algorithm
at small values of m. We fix $b=0.015$ and $mbN=6$
making the systematical error constant. The results are shown on
Fig.5, the fit of these data gives $\ti ({\cal R}_b)\approx
\frac{C}{m^\alpha}, C\approx 3.3, \alpha\approx 2$.

\section{Discussion.}

We performed the first simulations for Slavnov's algorithm
on 1D lattice for free fermions. It was shown that correct and accurate
results can be obtained with a reasonable size of lattice in auxiliary
dimension. We are going to extend this simulations to
larger lattices taking into account interaction with the gauge fields (in
progress).

\section{Acknowledgements.}

This research was supported by Russian Basic Research Fund under grants
96-01-00551, 96-02-17230a, INTAS-RFBR 95-0681 and by JSPS Program on
Japan-FSU scientists collaboration.

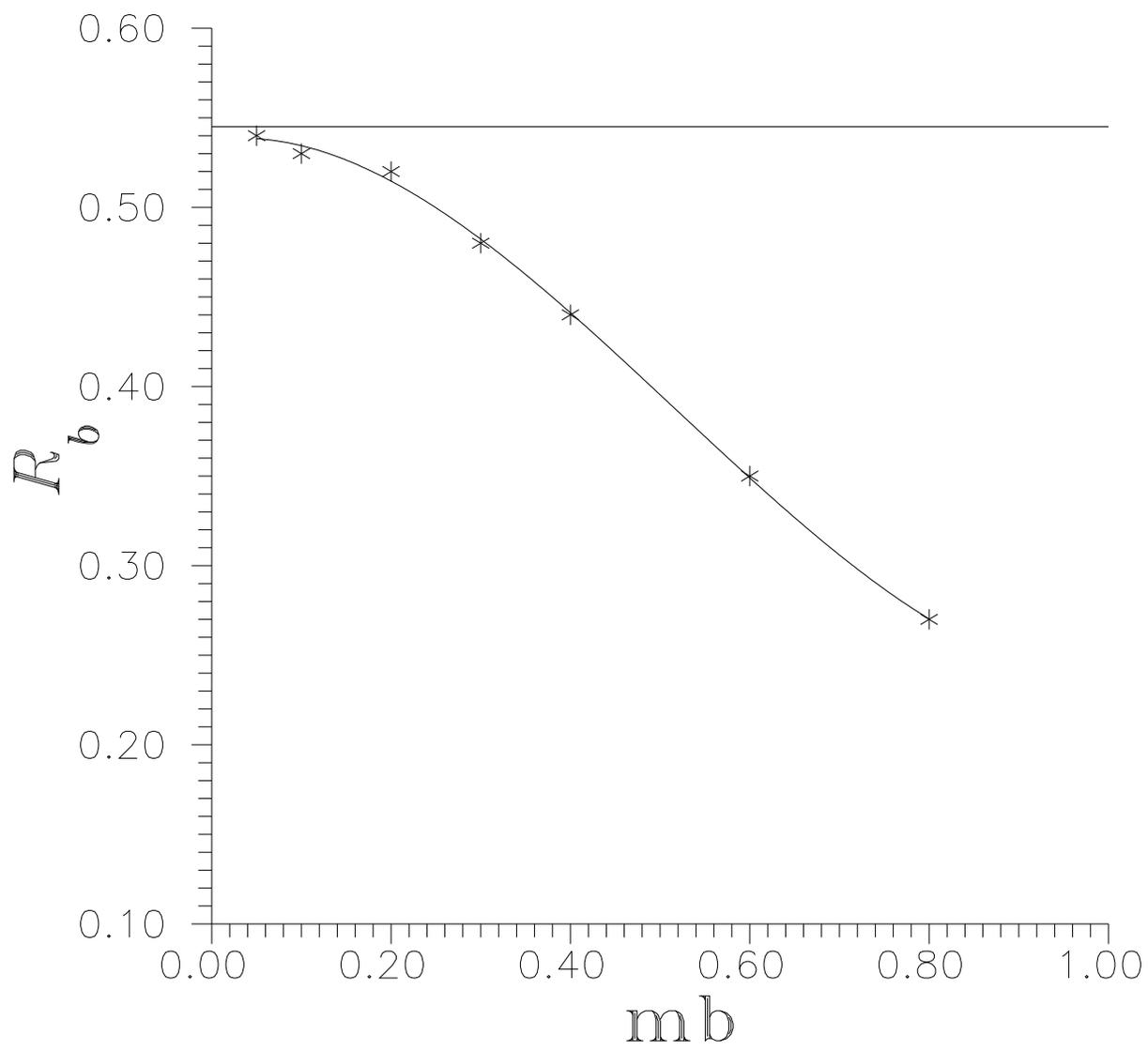
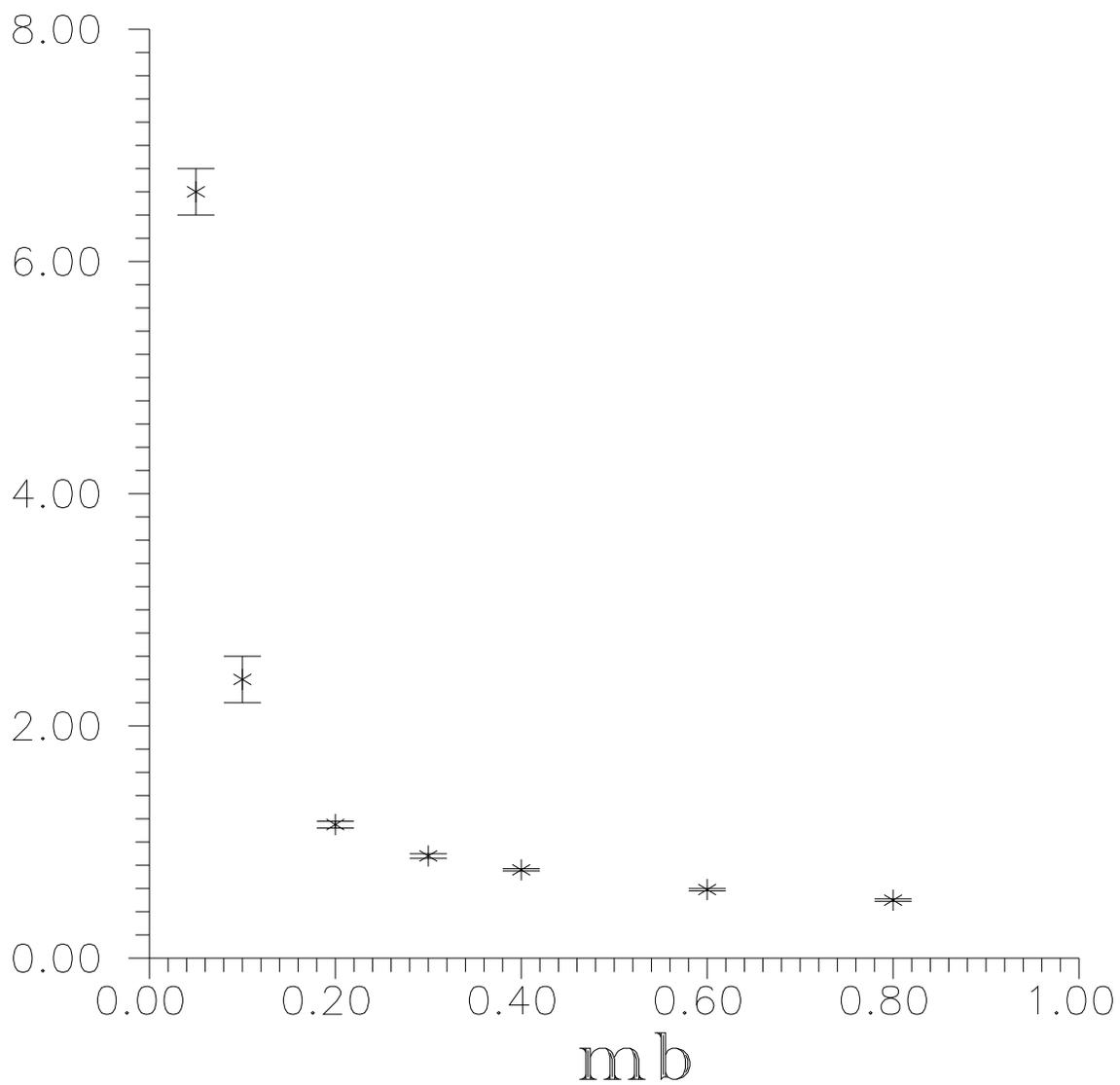
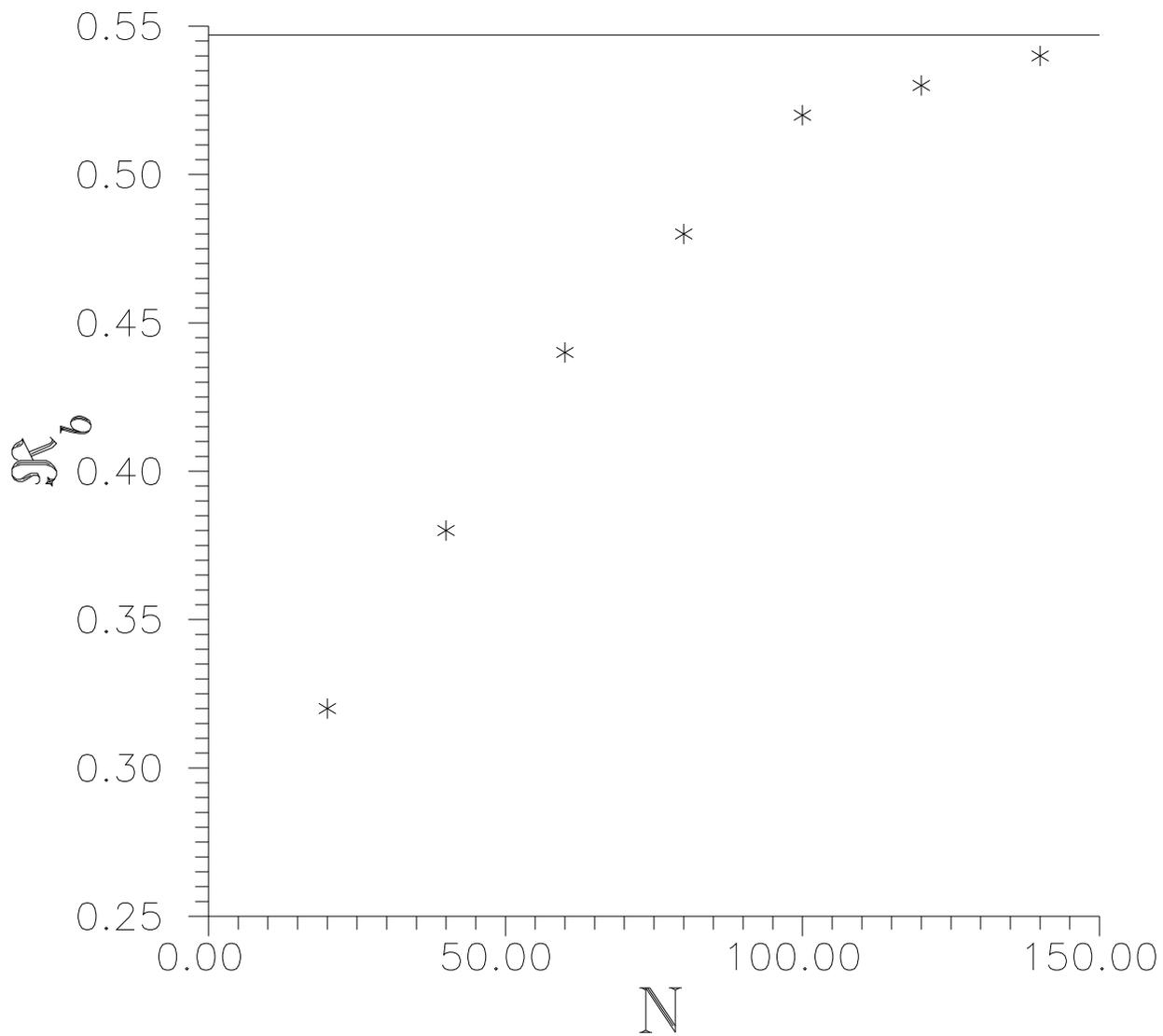
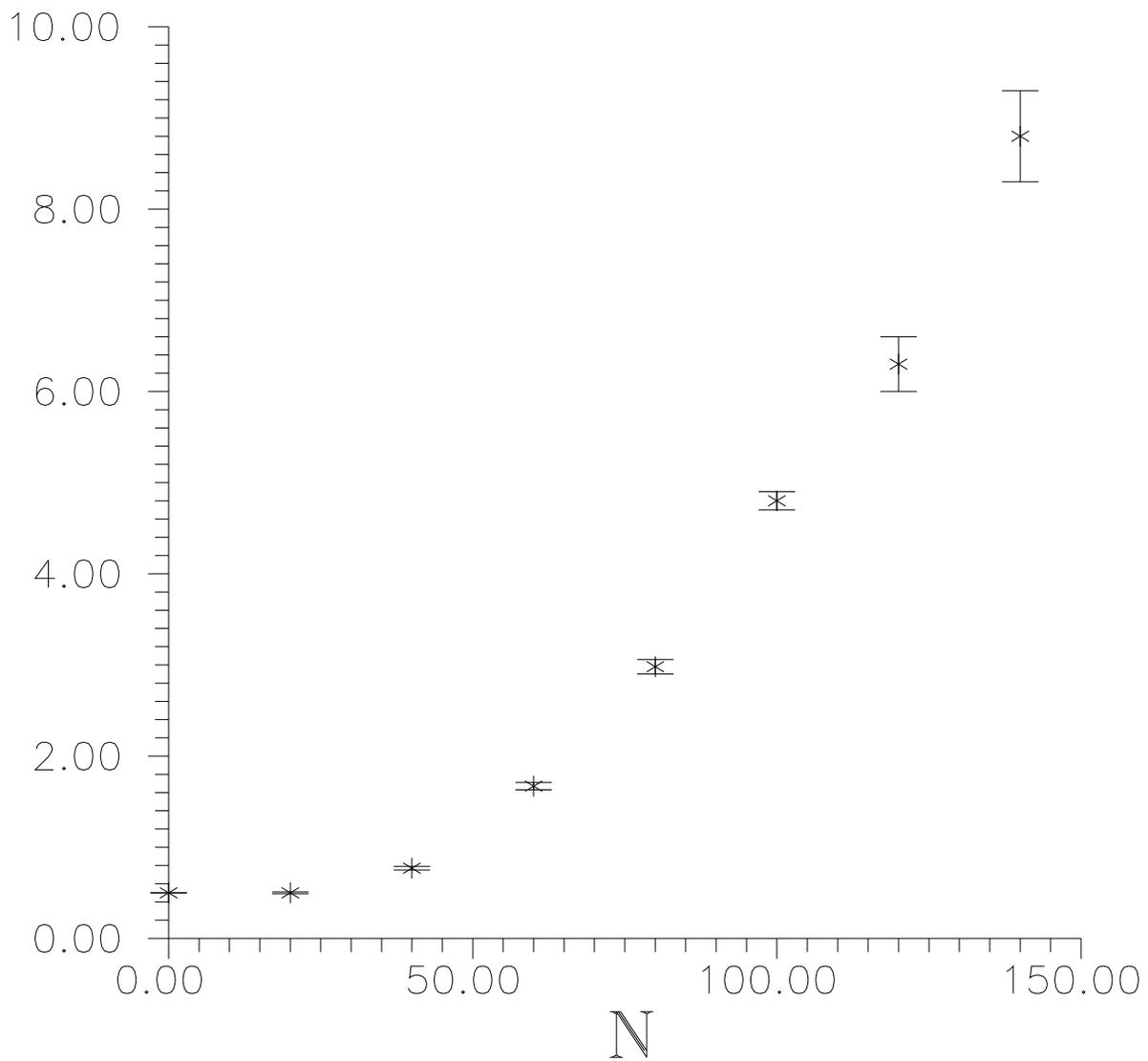
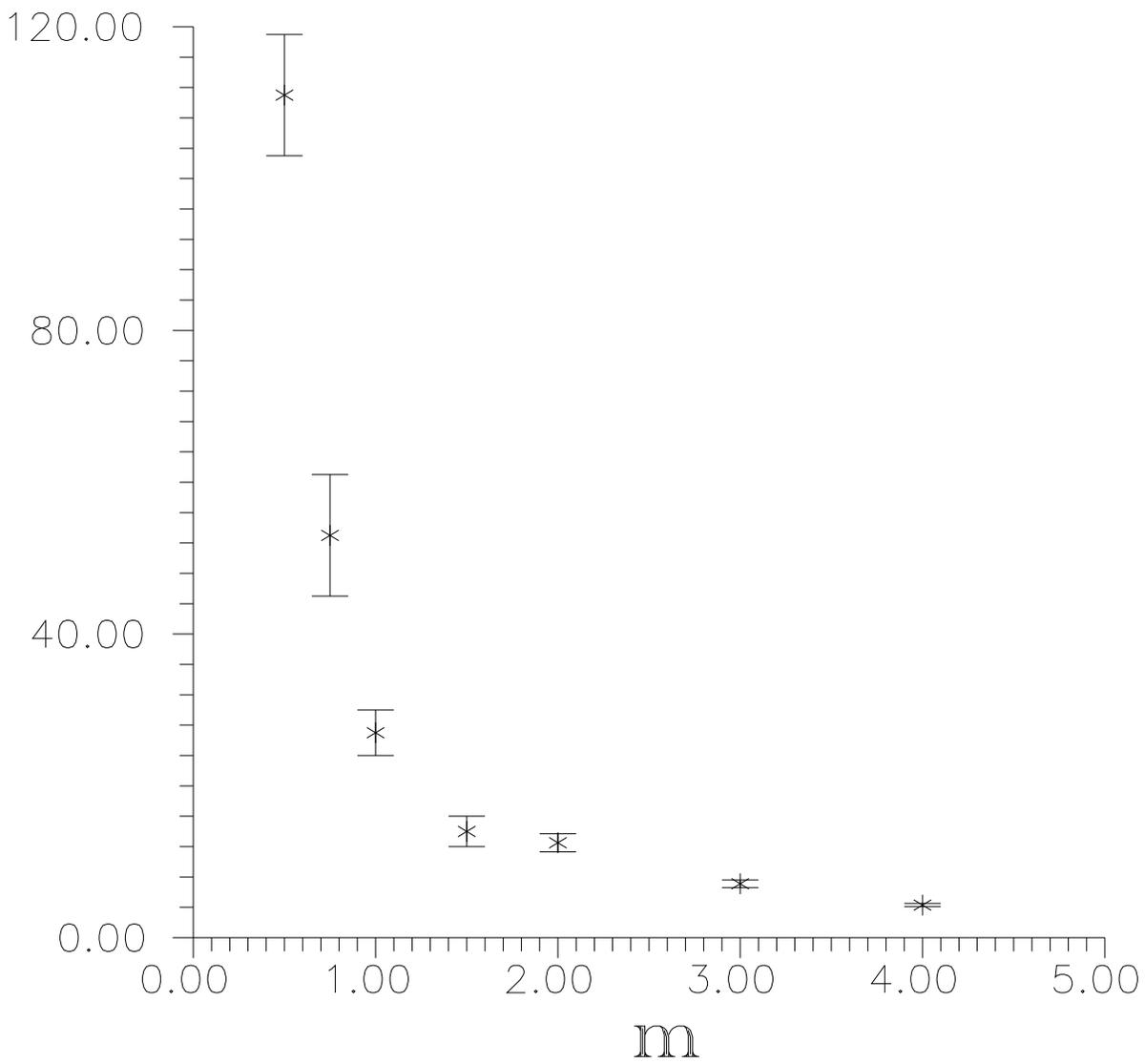
\end{document}